\documentclass[twocolumn]{aastex63}

\usepackage{amsmath}
\usepackage{amssymb}
\usepackage{graphicx}
\usepackage{xcolor}
\usepackage{multirow}

\newcommand{\Om}{\Omega_\mathrm{m}}
\newcommand{\Ob}{\Omega_\mathrm{b}}
\newcommand{\s}[1]{\sigma_{#1}}
\newcommand{\ee}[1]{\times10^{#1}}
\newcommand{\Msun}{\mathrm{M}_\odot}
\newcommand{\unit}[1]{\,\mathrm{#1}}

\shorttitle{Cosmology and baryons via deep learning from WL}
\shortauthors{Lu et~al.}

\begin{document}

\title{Simultaneously constraining cosmology and baryonic physics via deep learning from weak lensing}

\correspondingauthor{Tianhuan Lu}
\email{tl2854@columbia.edu}

\author[0000-0003-1040-2639]{Tianhuan Lu}
\affiliation{Department of Astronomy, Columbia University, New York, NY 10027, USA}

\author[0000-0003-3633-5403]{Zolt\'an Haiman}
\affiliation{Department of Astronomy, Columbia University, New York, NY 10027, USA}

\author[0000-0002-6267-716X]{Jos\'{e} Manuel Zorrilla Matilla}
\affiliation{Department of Astrophysical Sciences, Peyton Hall, Princeton University, Princeton, New Jersey 0854, USA}

\begin{abstract}
Ongoing and planned weak lensing (WL) surveys are becoming deep enough to contain information on angular scales down to a few arcmin. To fully extract information from these small scales, we must capture non-Gaussian features in the cosmological WL signal while accurately accounting for baryonic effects. In this work, we account for baryonic physics via a baryonic correction model that modifies the matter distribution in dark matter-only $N$-body simulations, mimicking the effects of galaxy formation and feedback.  We implement this model in a large suite of ray-tracing simulations, spanning a grid of cosmological models in $\Om-\s8$ space. We then develop a convolutional neural network (CNN) architecture to learn and constrain cosmological and baryonic parameters simultaneously from the simulated WL convergence maps. We find that in a Hyper-Suprime Cam (HSC)-like survey, our CNN achieves a 1.7$\times$ tighter constraint in $\Om-\s8$ space ($1\sigma$ area) than the power spectrum and 2.1$\times$ tighter than the peak counts, showing that the CNN can efficiently extract non-Gaussian cosmological information even while marginalizing over baryonic effects. When we combine our CNN with the power spectrum, the baryonic effects degrade the constraint in $\Om-\s8$ space by a factor of 2.4, compared to the much worse degradation by a factor of 4.7 or 3.7 from either method alone. 
\end{abstract}


\section{Introduction}
\label{sec:introduction}

When light travels from a distant galaxy to us, its path is bend by the gravitational influence of the matter in between. This causes the shapes of distant galaxies to be slightly distorted, in a phenomenon referred to as weak gravitational lensing (WL). By observing shapes of millions of galaxies over the sky, we can study the properties of the large-scale structure in our Universe \citep[see, e.g.][for reviews]{bartelmann2001,kilbinger2015}. In the framework of the Lambda Cold Dark Matter ($\Lambda$CDM) cosmology, WL signals are most sensitive to two of the cosmological parameters---the matter density $\Om$ and the amplitude of matter fluctuations $\s8$, precise measurements of which have been obtained by many recent analyses \citep{joudaki2016, kohlinger2017, hikage2019, hamana2020}. 

The depth of WL surveys nowadays has reached a galaxy number density of $\sim20\unit{arcmin^{-2}}$, and many surveys planned in the future are expected to have depths as high as $50\unit{arcmin^{-2}}$ \citep{aihara2018, laureijs2011, spergel2015, ivezic2019}. As a result, the information from WL signals will be available on very small scales (a few arcmin), which comes with two major challenges. 

First, WL signals are highly non-Gaussian on small scales. This poses a challenge to commonly used Gaussian statistics, such as the power spectrum and the two-point correlation function, as they only extract Gaussian information from the WL signal. To address this deficiency, many non-Gaussian summary statistics have been proposed and used, including three-point correlation functions \citep{takada2003, schneider2003, fu2014}, peak counts \citep{kratochvil2010,dietrich2010, liu2015}, and Minkowski functionals \citep{mecke1993, munshi2011, kratochvil2012, petri2013}.

Second, the small scale information is heavily mixed with the effects of many other astrophysical processes. These \emph{baryonic effects} are of great importance, due to the large uncertainty about how much they affect WL signals. One approach to minimizing the impact of these effects is to forego their modeling, and simply ignore the small scale information. For example, many WL studies have used power spectra with maximum multipoles $\ell$ of $\sim2000$, corresponding to angular scales above $10\unit{arcmin}$  \citep{kohlinger2016, kohlinger2017, hikage2019}. However, using a baryonic correction model \citep[BCM,][]{arico2020}, \citet[][hereafter LH21]{lu2021} have shown that increasing the maximum $\ell$ from 2000 to 12000 can improve the constraints in $\Om-\s8$ space by $\approx50\%$ (the area of $1\sigma$ credible contour) in an Hyper-Suprime Cam (HSC)-like survey ($A_\mathrm{survey}=1500\unit{deg^2}, n_\mathrm{g}=20\unit{arcmin^{-2}}$), after marginalizing over baryonic physics. They have also found that this marginalization over the uncertain BCM parameters (compared to fixing these parameters) degrade the constraints in $\Om-\s8$ space by factors of $4-5$ from either the power spectrum or the peak counts (counting the number density of peaks in a convergence map at difference convergence values). These results mean that modelling the baryonic effects is crucial in assessing the performance of summary statistics.

Convolutional neural networks (CNNs) offer a promising approach to the problems of extracting non-Gaussian information and marginalizing over baryonic effects. Instead of hand-crafting summary statistics, machines can learn the characteristics of the WL signals and automatically construct a statistic. Compared to fully connected neural networks, CNNs share weights (called kernels) across different parts of the input and can find patterns across different scales, typically from small scales to large scales. CNNs have achieved state-of-the-art results in processing image-like data, including image classification \citep{krizhevsky2012, he2016}, object detection \citep{ren2015, redmon2016}, and solving board games \citep{silver2018}. 

In the field of WL, several attempts have been made recently to use CNNs to infer cosmological parameters from WL convergence maps. \citet{gupta2018} trained a CNN on noiseless convergence maps and showed that the CNN yielded 5 times tighter constraints in $\Om-\s8$ space compared to the power spectrum. \citet{ribli2019} employed a different architecture of CNN and trained the networks on convergence maps with different levels shape noise, and they suggested that CNN can achieve $2.4–2.8$ times tighter constraints in an LSST-like survey. Using the KiDS-450 tomographic WL dataset, \citet{fluri2019} have found a 30\% improvement in the constraint on $S_8=\s8(\Om/0.3)^{0.5}$ by CNNs compared to the power spectrum.  \citet{matilla2020} have found that their CNN improves the constraints by 20\% compared to the combination of the power spectrum, peak counts, and Minkowski functionals. They also applied saliency methods to interpret CNNs on a map level, and found that the most relevant pixels are those with extreme values of the lensing convergence.
\citet{villaescusa2020} applied CNNs to Gaussian random fields and a toy model for baryons, to demonstrate that CNNs are able to recover all the information in the power spectrum and to infer cosmology, and that they are also able to learn and marginalize over toy baryonic effects.

In this study, we use a large suite of cosmological ray-tracing dark matter-only $N$-body simulations, augmented by a state-of-the art baryonic correction model, to explore the ability of CNNs to constrain both cosmology and baryonic physics. We train CNNs on WL convergence maps with baryons and compare their performance to that of power spectra and peak counts. 

This paper is organized as follows. In \S~\ref{sec:methods}, we introduce the generation of convergence maps (including the modelling of baryons), the architecture of our CNNs, and the inference of cosmology and baryon parameters. In \S~\ref{sec:results}, we show the prediction of parameters by the neural network and their derived posteriors in both cosmological and baryonic parameter space. In \S~\ref{sec:discussion}, we discuss how different aspects of the network architecture affect its performance, as well as the impact of shape noise. Finally, we summarize our main conclusions and the implications of our results in \S~\ref{sec:conclusions}.

\section{Methods}
\label{sec:methods}

\subsection{The baryonic correction model}

\begin{table*}[!ht]
\centering
\begin{tabular}{ccccc}
\hline
Parameter & Meaning & Fiducial value & Prior bound \\
\hline
$M_\mathrm{c}\ [h^{-1}\Msun]$ & characteristic halo mass for retaining half of the total gas & $3.3\ee{13}$ & $\left[5.9\ee{11},4.4\ee{15}\right]$ \\
$M_{1,0}\ [h^{-1}\Msun]$ & characteristic halo mass for a galaxy mass fraction of 0.023 & $8.63\ee{11}$ & $\left[9.3\ee{10},1.1\ee{13}\right]$ \\
$\eta$ & maximum distance of the ejected gas from the parent halo & 0.54 & $\left[0.12,2.7\right]$ \\
$\beta$ & logarithmic slope of the gas fraction {\it vs.} the halo mass & 0.12 & $\left[0.026,3.8\right]$ \\
\hline
\end{tabular}
\caption{Free parameters of the baryonic correction model}
\label{tab:baryon-priors}
\end{table*}

\begin{figure}[!ht]
\centering
\includegraphics[width=8cm]{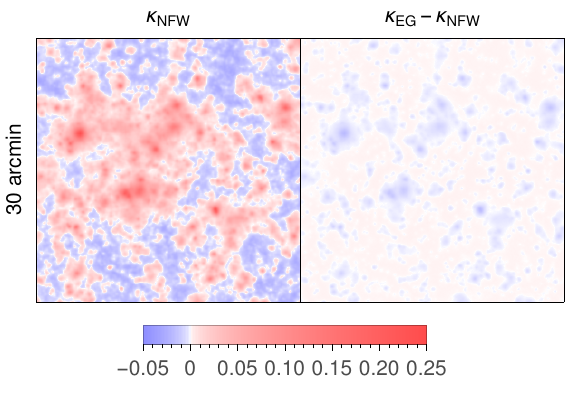}
\caption{The comparison of sample convergence ($\kappa$) maps using pure NFW halos ($\kappa_\mathrm{NFW}$) and using halos with the ejected gas components ($\kappa_\mathrm{EG}$). The ejection of gas from halos has the largest impact on our results; as illustrated in the right panel, this effect reduces $\kappa$ values at the highest peaks, mimicking a slight smoothing.}
%ZH3: looks nice. I suppose the "kappa_EG" map looked identical to the kappa_NFW, so you decided to show only their difference , and not kappa_EG by itself? I also added a 'take-away message' sentence above.
\label{fig:kappa-baryons}
\end{figure}

We employ the BCM to characterize the modification of the density distributions in and around each dark matter halo on a halo-by-halo basis. In the BCM, each halo after the modifications includes four components: the central galaxy (stars), bounded gas, ejected gas (due to active galactic nucleus feedback), and relaxed dark matter. The masses and profiles of these components are controlled by four free parameters: $M_\mathrm{c}$, $M_{1,0}$, $\eta$, and $\beta$. The meanings, fiducial values, and priors of these baryonic parameters can be found in Table~\ref{tab:baryon-priors}. 

For each halo in the dark matter-only $N$-body simulation, the masses of the baryonic components are calculated based on the halo mass and the redshift \citep[see][Appendix A]{arico2020}, which sums up to a fixed proportion ($\Ob=0.046$) of the halo mass. When the baryonic components are added to the halo, the same amount of mass is removed from the halo, so that the total mass is conserved. We show the effect of ejected gas as an example in Fig.~\ref{fig:kappa-baryons}---when we add the ejected gas component to the halos, whose profile is much wider than the NFW profile, the high peaks in the $\kappa$ map generally have lower $\kappa$ values. 

\subsection{Convergence maps with baryons}

The process of generating convergence maps is identical to that in LH21. We briefly summarize this process below and refer the reader to LH21 for a detailed description. 

The generation of convergence maps starts from a suite of 75 independent $N$-body simulations with different cosmologies (combinations of $\Om$ and $\s8$), where $(\Om=0.311,\s8=0.789)$ is chosen to be the fiducial cosmology. Each simulation has a linear box size of $240\,h^{-1}\unit{Mpc}$ (comoving) and contains ${512}^3$ dark matter particles. The simulations are run with the $N$-body code \textsc{gadget-2} \citep{springel2005}, and a series of snapshots are recorded between redshifts $0<z<1$ such that adjacent snapshots have a difference of $80\,h^{-1}\unit{Mpc}$ in comoving distance. The simulation snapshots are then post-processed to include baryonic effects. 

We use \textsc{rockstar} \citep{behroozi2012} to find all haloes with mass $>10^{12}\,\Msun$ and replace the particles in each halo by a spherically symmetric NFW profile with baryonic modifications determined by the chosen values of the four parameters. We note that the morphological differences between the haloes in the $N$-body simulations and spherical NFW haloes is not statistically significant, when compared to the uncertainties of the power spectrum and peak counts in an HSC-like survey (LH21).

The three-dimensional simulation box of each snapshot is then cut into three $80\,h^{-1}\unit{Mpc}$ thick slabs and each slab is projected into a two-dimensional lens plane.  We calculate a $4096\times4096$ potential map of each lens plane by solving a two-dimensional Poisson equation. Then, we ray-trace through a series of potential maps from $z=0$ to $z=1$ to obtain the total accumulated deflections, and to generate one realization of the corresponding convergence map. In this study, this process of ray-tracing is repeated 2048 times for each cosmology (increasing from 1280 in LH21). Each pseudo-independent realization has its unique way of randomizing the potential maps (rotating, flipping, and shifting the lens planes) and its choices of random baryonic parameters. The randomization are kept the same across all cosmologies, i.e., the $i$-th realization of every cosmology shares the same baryonic parameters and the same rotation, flip, and shift. The same set of baryonic parameters for all cosmologies makes the interpolation of the statistics between cosmologies easier. 

The raw noiseless convergence maps have a resolution of $2048\times2048$ pixels, and a field of view of $3.5\times3.5\unit{deg^2}$ (corresponding to the linear size of the box at the highest redshift, $z=1$).
We add shape noise to each pixel of the map following a Gaussian distribution  ($\mu=0,\sigma=\sigma_\epsilon/\sqrt{2n_\mathrm{gal}A_\mathrm{pixel}}$), where $\sigma_\epsilon=0.4$ is the mean intrinsic ellipticity of galaxies and $A_\mathrm{pix}$ the area of the pixel. Finally, we smooth the map with a $1\unit{arcmin}$ Gaussian filter. 

\subsection{Convolutional neural network}

\subsubsection{Architecture}

\begin{figure}[!ht]
\centering
\includegraphics[width=3.8cm]{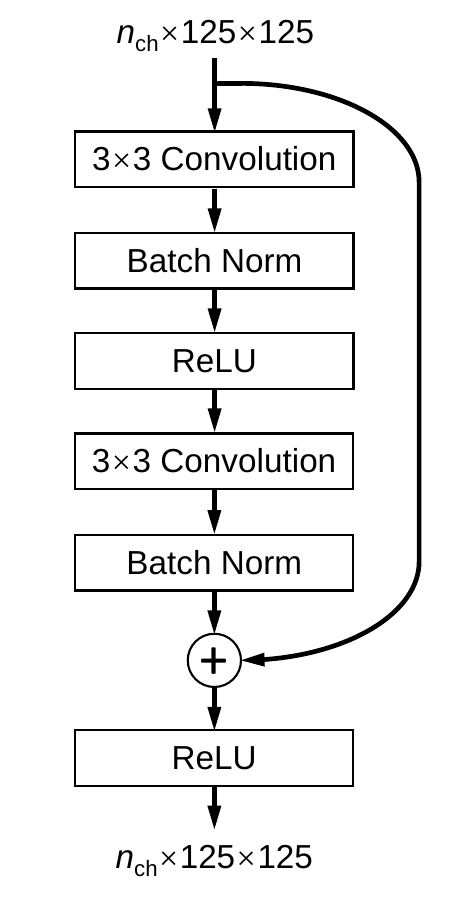}
\caption{The architecture of the residual block. The $3\times3$ convolution layers have the same number of output and input channels. The input image is padded by one pixel on each side before convolution, so that its dimension is preserved.}
\label{fig:resblock}
\end{figure}

\begin{table}[!ht]
\centering
\begin{tabular}{cccc}
\hline
Layer & Kernel size & Stride & Output dimensions \\ \hline
(Input)     &            &   & $(1\times512\times512)$               \\
Convolution & $5\times5$ & 2 & $(n_\mathrm{ch}/2)\times254\times254$ \\
Convolution & $5\times5$ & 2 & $n_\mathrm{ch}\times125\times125$     \\

Residual block &  &  & \\
$\vdots$       &  &  & \\
Residual block &  &  & $n_\mathrm{ch}\times125\times125$ \\

Pooling     & $2\times2$ & 2   & $     n_\mathrm{ch} \times62\times62$ \\
Convolution & $3\times3$ & 1   & $( 2\,n_\mathrm{ch})\times60\times60$ \\
Pooling     & $2\times2$ & 2   & $( 2\,n_\mathrm{ch})\times30\times30$ \\
Convolution & $3\times3$ & 1   & $( 4\,n_\mathrm{ch})\times28\times28$ \\
Pooling     & $2\times2$ & 2   & $( 4\,n_\mathrm{ch})\times14\times14$ \\
Convolution & $3\times3$ & 1   & $( 8\,n_\mathrm{ch})\times12\times12$ \\
Pooling     & $2\times2$ & 2   & $( 8\,n_\mathrm{ch})\times 6\times 6$ \\
Convolution & $3\times3$ & 1   & $(16\,n_\mathrm{ch})\times 4\times 4$ \\
Pooling     & $4\times4$ & $-$ & $(16\,n_\mathrm{ch})\times 1\times 1$ \\
Linear      & $-$        & $-$ & $256$                                 \\
ReLU        & $-$        & $-$ & $256$                                 \\
Linear      & $-$        & $-$ & $6$                                   \\ \hline

\end{tabular}
\caption{The architecture of the CNNs. Here $n_\mathrm{ch}$ (number of channels) and $n_\mathrm{layer}$ (number of residual blocks) are hyper-parameters. Convolution layers, except those in the residual blocks, are implicitly followed by batch normalization layers and then ReLU layers. See text for a detailed description.} 
\label{tab:architecture}
\end{table}

In this study, our CNNs take $512\times512$ convergence maps as the inputs, which are down-sampled from the $2048\times2048$ map by averaging $4\times4$ patches. Accordingly, the resolution of the maps is reduced from $9.8\unit{pixel/arcmin}$ to $2.4\unit{pixel/arcmin}$. This down-sampling step loses very little information because the maps are already smoothed on a scale of $1\unit{arcmin}$, but it brings the advantage of significantly speeding up the training and evaluation of the CNNs.

The CNNs consist of three parts in series. The first part of the network transforms the input (a $512\times512$ convergence map) into an $n_\mathrm{ch}\times125\times125$ image (meaning a $125\times125$ image with $n_\mathrm{ch}$ channels) through two convolutions. The convolutions are applied to every other pixel on the images, so the width and height of the output are approximately half of the input, and each channel in an output image corresponds to a kernel in the convolution. The second part consists of a sequence of $n_\mathrm{layer}$ residual blocks \citep{he2016}, where the operation of each block preserves the dimension of the image. The structure of the residual blocks is shown in Fig.\ref{fig:resblock}. A residual block contains a skip connection, which provides an alternative path for the input image to bypass the convolutions and enable the network to learn effectively even if it is very deep (with many residual blocks). The third part transforms the image into six numbers corresponding to the two cosmological and four baryonic parameters, again through a number of convolutions. The structure of our CNNs is shown in Table~\ref{tab:architecture}. Note that $n_\mathrm{ch}$ and $n_\mathrm{layer}$ are considered two hyper-parameters of this architecture, i.e., different networks can be constructed by choosing different values of $n_\mathrm{ch}$ and $n_\mathrm{layer}$. For the results in Section~\ref{sec:results}, we employ a network with $n_\mathrm{layer}=10$ and $n_\mathrm{ch}=64$ (called the \emph{main network}), and the results with other choices are discussed in Section~\ref{sec:alternative-arch}. 

We adopt the rectified linear unit (ReLU, defined as $f(x)=\max\{x, 0\}$) activation functions for our networks, and before most of the activation layers, we insert batch normalization layers (re-centering and re-scaling the inputs) to improve training speed and network stability \citep{ioffe2015}. The loss of our network is chosen to be the mean-squared error (MSE) between the output $y_i$ ($i=1,2,\dotsc,6$) and the true parameters after re-scaling to $[0,1]$ on a logarithmic scale (called the \emph{targets}) used for generating the input map $\theta_i$:
\begin{equation}
L=\frac{1}{6}\sum_{i=1}^{6}{(y_i-\theta_i)^2}.
\end{equation}
For $\Om$ and $\s8$, we take their logarithms to be the targets. For the four baryonic parameters, we take their logarithms and re-scale them from the priors to $[0,1]$ as the targets. By doing these transformations, we make the ranges of all the targets have widths of about unity, and they also have similar standard deviations (all of the parameters contribute to the loss function roughly equally). 

\subsubsection{Training and testing}
\label{sec:training}

\begin{figure*}[!ht]
\centering
\includegraphics[width=14cm]{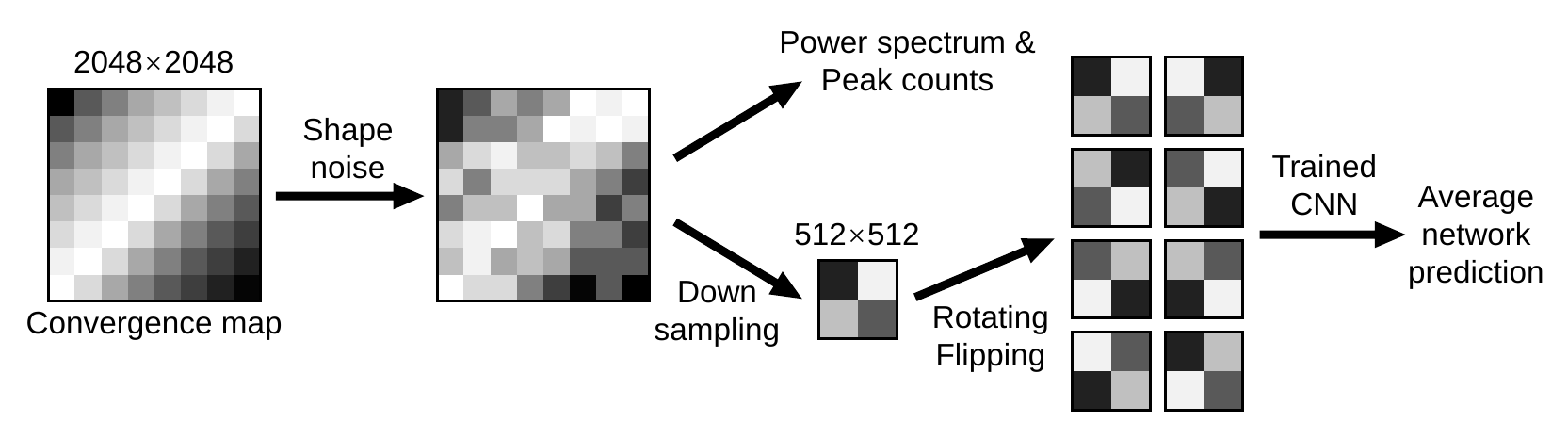}
\caption{An illustration of how we obtain the summary statistics, including the predictions by the trained CNN, from each convergence map in the test set. The results are then used for parameter inference as explained in the text.}
\label{fig:testing}
\end{figure*}

To make a fair evaluation of the performance of the network, we need to split all convergence maps into two sets: \emph{the training set}, from which the network learns, and \emph{the test set}, for which the network makes predictions, which we then use to calculate posteriors on the parameters. In addition, the information that is shared between the two sets should be minimized. In this study, we put the first 1024 maps in each cosmology into the training set and the remaining 1024 maps into the test set. Since we keep the randomization of the potential planes the same when generating the $i$-th convergence map in all cosmologies, this method of splitting ensures that the network is tested on light-cones that have never been seen during its training. 

We use the stochastic gradient descent optimizer (momentum parameter $\alpha=0.9$) to train the network for 30 epochs, and each batch contains 75 sample maps (one from each cosmology). We set the initial learning rate to $10^{-2}$, lower it to $10^{-3}$ after 20 epochs, and lower it again to $10^{-4}$ after 5 more epochs. Since the cosmological and baryonic parameters are invariant to the rotation and flip of the maps, we augment each map by randomly rotating and flipping it during the training. We also add random shape noise to the maps, so that the same map looks different to the network every time it is sampled. The training runs were performed on the NSF TACC Frontera cluster, with each run typically taking four hours using four Quadro RTX 5000 GPUs, although this run time varies with different network architectures.

After training, we use the network to make predictions of cosmological and baryonic parameters. For each map in the test set, we add the same amount of shape noise (but with an independent realization) as in the training, 
down-sample it to the network's input resolution ($512\times512$), and make eight copies of the map by rotating and flipping. We apply the trained network to each of these, and take the average of the eight predictions. An illustration of this process is shown in Fig.~\ref{fig:testing}.

\subsection{Lensing power spectrum and peak counts}

In addition to CNNs, we evaluate two commonly used statistics, the power spectrum and the peak counts, on the convergence maps. We compare their ability to constrain the parameters to that of the CNNs, and also use them in conjunction with the CNNs. We calculate the power spectra from the Fourier transforms of the maps, taking the values in 18 equally spaced bins within the multipole range $100<\ell<12000$ on a logarithmic scale. The upper limit of $\ell=12,000$ corresponds to an angular scale of $\sim1\unit{arcmin}$, the scale on which we smooth the convergence maps. We take the logarithm of the power spectrum in each bin to be the observable for parameter inference. The peak counts of each map are defined as the number of peaks (i.e., pixels whose convergence $\kappa$ are higher than the convergence of their eight neighbours) per $\deg$ in each of 18 $\kappa$ bins, which are equally spaced within $-0.03<\kappa<0.15$. Since the mean and variance of the number of peaks in each bin vary by orders of magnitude, we take the logarithm of each peak count to be the observable.

\subsection{Parameter inference}
\label{sec:parameter-inference}

For the purpose of parameter inference, we note that the output of a neural network can be regarded the same as a pre-specified statistic, such as the power spectrum or the peak counts. That is, we consider the predictions given by the network to be the values of a statistic, regardless of their connection to the underlying parameters. This method has been used in our previous studies \citep[e.g.][]{gupta2018, ribli2019, matilla2020}. Therefore, a \emph{statistic} mentioned below can refer to the output prediction by a network or its combination with other statistics.

Given a summary statistic with $d$ observables, we model the likelihood of observing $\mathbf{y}$ as a multidimensional Gaussian distribution:
\begin{gather}
p(\mathbf{y}|\theta)\propto\frac{1}{\sqrt{\det\mathbf{C}}}\exp\left(-\frac{1}{2}\Delta\mathbf{y}^\mathrm{T}\widehat{\mathbf{C}^{-1}}\Delta\mathbf{y}\right), \label{eqn: delta-y}\\
\Delta \mathbf{y}=\mathbf{y}-\mathbf{y}(\theta), \\
\widehat{\mathbf{C}^{-1}}=\frac{N-d-2}{N-1}\mathbf{C}^{-1}, \\
\mathbf{C}=\left(\frac{A_\mathrm{survey}}{3.5\times3.5\unit{deg^2}}\right)^{-1}\mathbf{C}_\mathrm{sim}(\Om,\s8),
\end{gather}
where $\theta$ denotes the underlying cosmological and baryonic parameters, $N$ the number of realizations to estimate the covariances $\mathbf{C}$, and $A_\mathrm{survey}$ the area of a hypothetical mock survey. Following LH21, the factor $(N-d-2)/(N-1)$ is included to make the estimation of the precision matrix $\widehat{\mathbf{C}^{-1}}$ unbiased \citep{hartlap2007}. We remove the randomness caused by the sampling of shape noise by taking each $\mathbf{y}(\theta)$ to be the average value of the observables calculated from 36 independent shape noise field realizations. 

To extend the likelihood from the 75 discrete sampled cosmologies to the entire $\Om-\s8$ space, we interpolate the expected value $\mathbf{y}(\theta)$ in two steps: (1) we fit a second-degree polynomial on the four baryonic parameters for each cosmology, and (2) the values between the cosmologies are interpolated linearly with Delaunay triangulation. When compared to fitting a polynomial globally (as is used in LH21), this method improves the estimation of distributions near the edge of the $\Om-\s8$ space, while it does not differ from LH21 near the fiducial cosmology. As for covariances, it has been found that their dependencies on baryonic parameters are much weaker than those on cosmological parameters over the ranges of posteriors (LH21). Therefore, we assume the covariances to be constant across different baryonic parameters following LH21; we estimate the covariances for each cosmology and interpolate them linearly with Delaunay triangulation.
%ZH3: last sentence is truncated....

We use Bayes' theorem to estimate the posterior distribution of $\theta$, given a mock observation at the fiducial parameters $\mathbf{y}(\theta_0)$. We take log-uniform priors on all parameters: $0.2<\Om<0.6$, $0.4<\s8<1.1$, and the range of baryonic parameters are shown in Table~\ref{tab:baryon-priors}. We sample the posterior with a Monte Carlo Markov chain of $10^6$ steps (identical to LH21), which is sufficient for the chains to converge.

To validate our modelling of the likelihoods of the CNN's predictions as multidimensional Gaussian distributions, we generate 128 convergence map realizations at the fiducial cosmological and baryonic parameters and test the normality of CNN's predictions (using our main network) on those maps with $n_\mathrm{g}=20\unit{arcmin^{-2}}$. Mardia's test \citep{mardia1970} for kurtosis and skewness gives $p$-values of 0.48 and 0.15, and the BHEP test \citep{baringhaus1988} gives a $p$-value of 0.64, none of which rejects the hypothesis that the predictions are normally distributed. A similar conclusion has also been found by \cite{gupta2018} for cosmological parameters only.

\subsection{Evaluation metrics: $S_\mathrm{full}$ and $S_\mathrm{fid}$}

Similarly to LH21, we measure the level of constraining ability quantitatively through the area $S$ enclosed by the credible contours containing 68\% of the total likelihood (hereafter ``$1\sigma$ area'') in the $\Om-\s8$ plane.  $S_\mathrm{full}$ refers to this area with all baryonic parameters marginalized over, and $S_\mathrm{fid}$ is the area in the conditional distribution with the baryonic parameters fixed at their fiducial values. Additionally, we calculate the ratio of these two areas $S_\mathrm{full}/S_\mathrm{fid}$ as an indicator of the degradation caused by the uncertainties in the baryonic parameters.

Here, we note a complication when the concept of $S_\mathrm{fid}$ is applied to our CNN. There are two different ways of fixing baryonic parameters: (1) training the network to predict both cosmological and baryonic parameters and fixing the baryonic parameters within the posterior distribution, and (2) training the network to predict cosmological parameters only with a set of maps with fixed baryonic parameters. We take the former approach because it reuses the networks that we have already trained, and it does not require additional maps to be generated. Using the same network between $S_\mathrm{full}$ and $S_\mathrm{fid}$ makes the comparison between them more direct and makes the evaluations consistent across all statistical methods. Additionally, we expect that fixing baryonic parameters during the training is effectively the same as not considering baryonic effects at all in terms of the $1\sigma$ area, because only $\Om$ and $\s8$ are causing changes to the maps in both cases. This case has already been investigated thoroughly in prior work (e.g., by \citealt{ribli2019}).

\section{Results}
\label{sec:results}

\subsection{Predictions by the CNN}

\begin{figure*}[!ht]
\centering
\includegraphics[width=12.5cm]{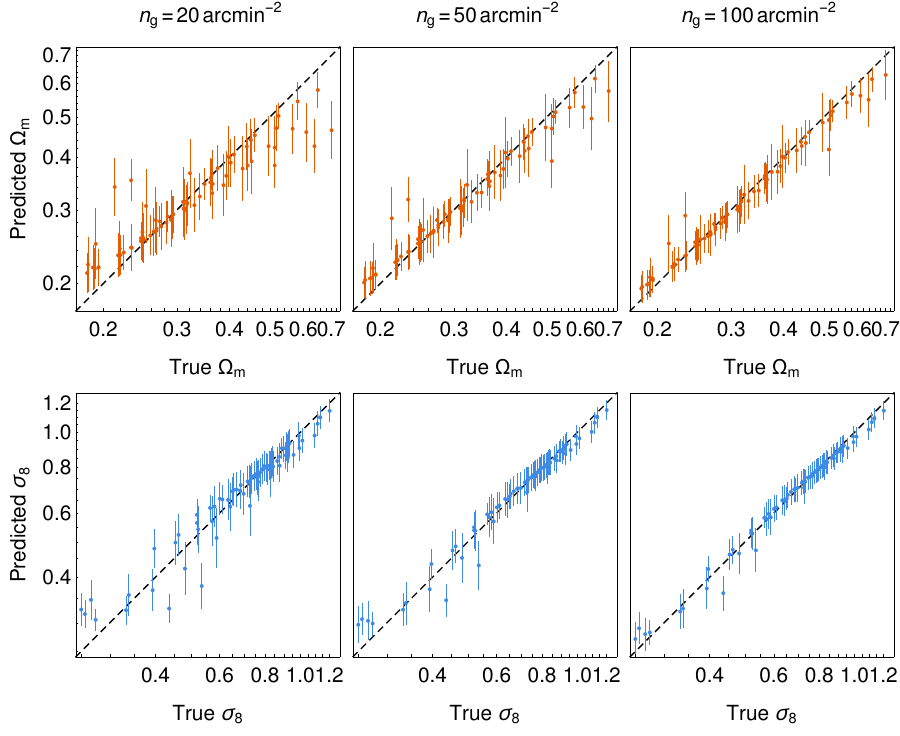}
\caption{The values of $\Om$ and $\s8$ predicted by the networks, trained with and tested on three shape noise levels (with noise decreasing to the right). Each point in the panels represent the average over the 1,024 predictions within the same cosmology, with the baryonic parameters spanning the full range of their priors. The error bars correspond to the standard deviation over these realizations. The dashed lines mark the predictions matching the true input values.}
\label{fig:predictions-cosmological}
\end{figure*}

\begin{figure*}[!ht]
\centering
\includegraphics[width=16.0cm]{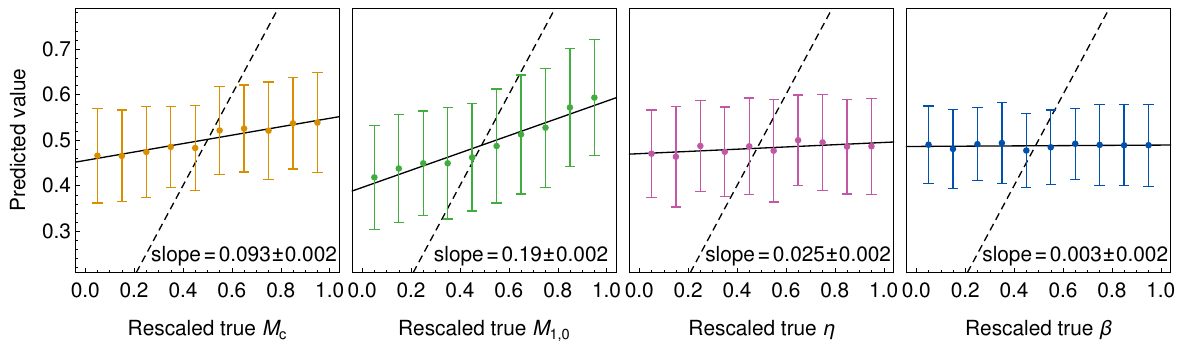}
\caption{The baryonic parameters predicted by the network in the fiducial cosmology at $n_\mathrm{g}=50\unit{arcmin^{-2}}$. The parameter values are re-scaled on a logarithmic scale from their priors to $\left[0,1\right]$. The solid line in each panel shows the best linear fit, with the slope indicated in the lower right corner. The dashed lines mark the predictions matching the true input values.}
\label{fig:predictions-baryonic}
\end{figure*}

In this section, we inspect the performance of our main network ($n_\mathrm{layer}=10,n_\mathrm{ch}=64$), by looking at its parameter predictions. The network is trained three times, each from a set of maps with a noise level of 20, 50, or $100\unit{arcmin^{-2}}$. In Fig.~\ref{fig:predictions-cosmological}, we show the predictions of cosmological parameters against the true values (i.e. the ``ground truth'', used in generating the maps). Each data point corresponds to the average over the 1,024 predictions for the same input cosmology, with the baryonic parameters ranging across the whole prior.  The error bars show the standard deviation over these 1,024 realizations. We find two main features from Fig.~\ref{fig:predictions-cosmological}. First, as the shape noise decreases, the spread of predictions within each cosmology generally decreases, and the mean predictions get closer to the true values. This shows that our network indeed learns the correlation between the maps and the cosmological parameters---when more information is available, it can make both more accurate and more precise predictions. 
Second, the predictions tend to be lower than the true value at high $\Om$, and they are higher at low $\Om$ (this also applies to $\s8$ but is less noticeable). We note that this is a common behavior of neural networks. The reason is that neural networks learn the prior distribution of the true values, and when the information is insufficient, predicting towards the center of the prior is beneficial in reducing the loss. Nevertheless, these biased predictions will not bias the estimation of posteriors because they are not directly used (see below). These two features were also found by \citet{gupta2018} and \citet{ribli2019} with different network architectures. We also find that the spreads are more even (i.e., the error bars in Fig~\ref{fig:predictions-cosmological} are of similar sizes) across the cosmologies in our result compared to \citet{ribli2019}, possibly due to our training including additional baryonic parameters. 

In Fig.~\ref{fig:predictions-baryonic}, we show the predictions of each baryonic parameter (with the other baryonic parameters being free) by the network in the fiducial cosmology and $n_\mathrm{g}=50\unit{arcmin^{-2}}$. It is apparent that our network learns $M_\mathrm{c}$ and $M_\mathrm{1,0}$ much better than $\eta$ and $\beta$. Here, we quantify the level of learning by fitting a straight line to the predictions. We find that the slope of $\eta$ is only $8\sigma$ away from zero, and the slope of $\beta$ is consistent with zero. A slope of zero means that the network is unable to learn the relationship between the maps and the parameters -- it keeps the loss low by predicting random values close the center of the prior. 

\subsection{Posterior distributions}

\begin{figure*}[!ht]
\centering
\includegraphics[width=17.5cm]{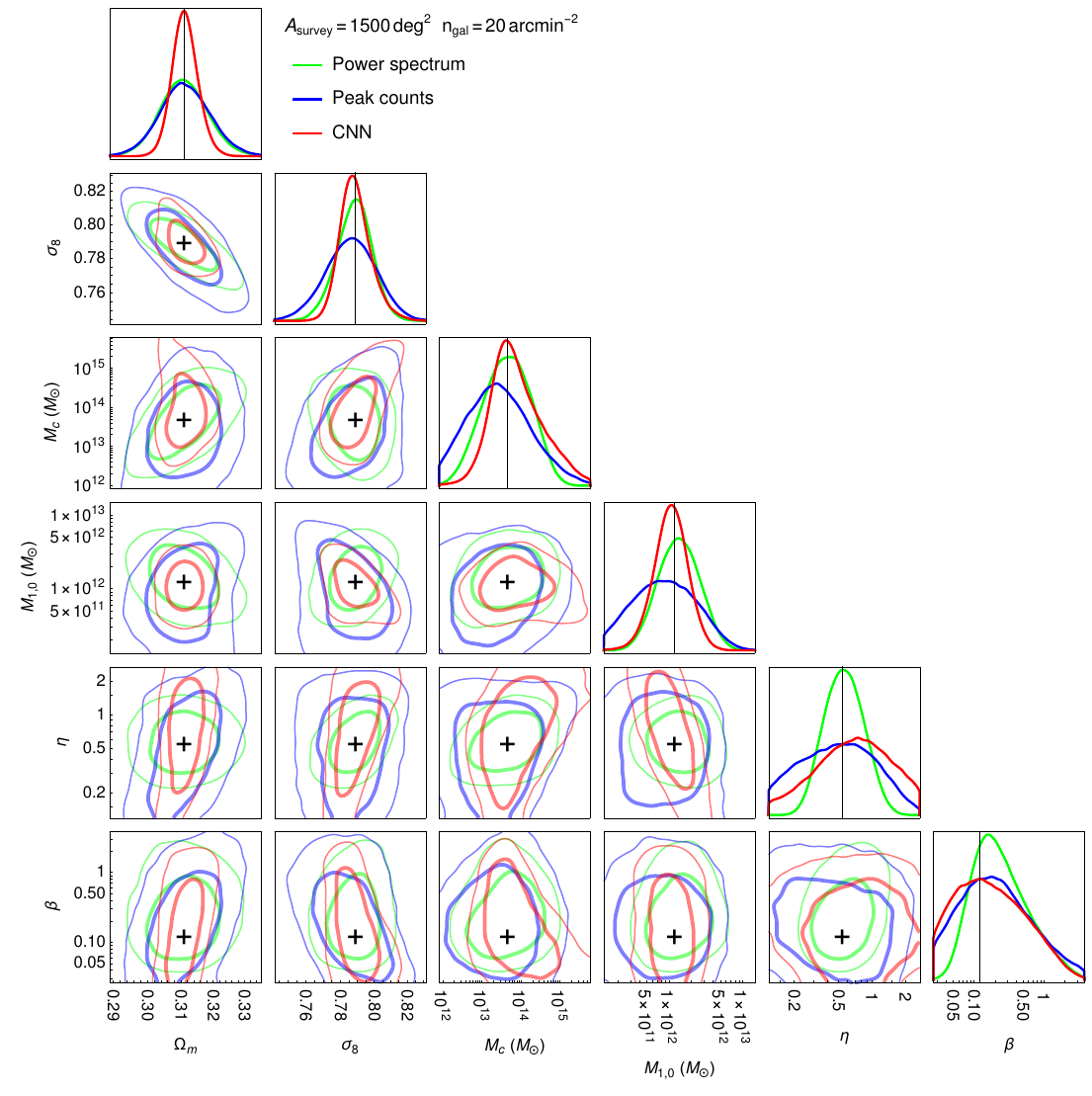}
\caption{The posteriors of cosmological and baryonic parameters in an HSC-like survey area and galaxy density. The thick and thin contours show the $1\sigma$ (68\%) and $2\sigma$ (95\%) credible region respectively, and the black crosses and lines show the fiducial values of the parameters.
The green/blue/red contours correspond to posteriors derived from the power spectrum, peak counts, and the fiducial neural network, respectively. The neural network outperforms both statistics.}
\label{fig:cornerplot-20}
\end{figure*}

\begin{table*}[!ht]
\centering
\begin{tabular}{ccccccc}
\hline
\multirow{2}{*}{Methods} &  \multicolumn{3}{c}{$\Om-\s8$}  &  \multicolumn{3}{c}{$M_{1,0}-\eta$} \\
& $S_\mathrm{full}/10^{-4}$ & $S_\mathrm{fid}/10^{-4}$ & $S_\mathrm{full}/S_\mathrm{fid}$ & $S_\mathrm{full}/10^{-2}$ & $S_\mathrm{fid}/10^{-2}$ & $S_\mathrm{full}/S_\mathrm{fid}$ \\ \hline
Power spectrum                           &
  3.45 & 0.93 & 3.71 & 10.4 & 3.6 & 2.88 \\
Peak counts                              &
  5.89 & 0.94 & 6.28 & 30.6 & 7.3 & 4.16 \\
CNN                                      &
  2.08 & 0.44 & 4.70 & 13.0 & 3.7 & 3.48 \\
CNN$+$Power spectrum (L)                 &
  1.27 & 0.44 & 2.91 &  7.1 & 2.6 & 2.69 \\
CNN$+$Power spectrum (M)                 &
  1.11 & 0.42 & 2.61 &  6.9 & 2.8 & 2.41 \\
CNN$+$Power spectrum (S)                 &
  1.74 & 0.41 & 4.23 &  9.7 & 3.0 & 3.26 \\
CNN$+$Power spectrum (L,M)               &
  1.01 & 0.42 & 2.39 &  5.2 & 2.3 & 2.24 \\
CNN$+$Power spectrum (full)              &
  0.96 & 0.40 & 2.41 &  4.6 & 2.1 & 2.24 \\ \hline
\end{tabular}
\caption{The area of the 68\% posterior distributions in $\Om-\s8$ or $M_{1,0}-\eta$ space in an HSC-like survey, marginalized over the other four parameters ($S_\mathrm{full}$) or with the other four parameters fixed at their fiducial values ($S_\mathrm{fid}$). In the case of $M_{1,0}-\eta$, the other four parameters consist of two cosmological and two baryonic. The last five rows show joint constraints from our fiducial CNN combined with the power spectrum at different scales: $100<\ell<400$ (L), $400<\ell<2000$ (M), and $2000<\ell<12000$ (S).
}
\label{tab:posterior-area}
\end{table*}

In Fig.~\ref{fig:cornerplot-20}, we show the posterior distribution of cosmological and baryonic parameters in an HSC-like survey ($A_\mathrm{survey}=1500\unit{deg^2}, n_\mathrm{g}=20\unit{arcmin^{-2}}$) with MCMC sampling as described in \S~\ref{sec:parameter-inference}. In terms of constraining cosmological parameters, the performance of our CNN is better than the power spectrum and the peak counts, which is shown in the $\Om-\s8$ panel of Fig.~\ref{fig:cornerplot-20}. We list the $1\sigma$ areas and $S_\mathrm{full}/S_\mathrm{fid}$ derived using different methods in Table~\ref{tab:posterior-area}. We find that when the baryonic parameters are fixed, our CNN achieves a $2.10$ times lower $S_\mathrm{fid}$ than the power spectrum, which proves that our CNN can efficiently extract non-Gaussian information from WL maps. Similar levels of improvements have been found in previous studies (with different networks): \citet{ribli2019} have found this ratio to be $2.4-2.8$ at a noise level of $30\unit{arcmin^{-2}}$, and \citet{fluri2018} have found it to be $\sim2.5$ at a noise level of $15\unit{arcmin^{-2}}$ and a map smoothing scale of $1.17\unit{arcmin}$. On the other hand, when we marginalize over baryonic parameters, the advantage of the CNN over the power spectrum narrows to a factor of 1.66, meaning that baryons have a greater negative impact on the CNN (this can also been seen from the CNN having a higher $S_\mathrm{full}/S_\mathrm{fid}$).

In terms of constraining baryonic parameters, our CNN does not outperform the power spectrum and the peak counts overall. While our CNN's constraints on $M_\mathrm{c}$ and $M_{1,0}$ is tighter compared to the other two methods, the marginal posterior distributions of $\eta$ and $\beta$ by the network only weakly prefer the fiducial baryonic parameters. This result is consistent with our finding with the directed predictions by the network. In this study, we pick $M_{1,0}$ and $\eta$ and define $S_\mathrm{full}$ and $S_\mathrm{fid}$ similar to $\Om$ and $\s8$ to quantify the ability of these methods in constraining baryonic physics. Our reason for picking these two parameters is that $M_{1,0}$ is relatively weakly constrained by external observations (e.g., \citealt{behroozi2013} suggests $M_{1,0}=(2-10)\ee{11}h^{-1}\Msun$), while $\eta$ has the greatest impact on WL signals among all baryonic parameters (LH21). The results are shown in Table~\ref{tab:posterior-area}. We find that our CNN and the power spectrum perform similarly when all other parameters are fixed, and the power spectrum is slighter better when marginalized over the other parameters (10.4 versus 13.0). The peak counts perform worse than the other two methods by a factor of around 2 in either case. 

\section{Discussion}
\label{sec:discussion}

\subsection{Combining CNN predictions with the power spectrum}
\label{sec:combine-cnn-ps}

From the fact that the power spectrum constrains $\eta$ and $\beta$ better than our CNN, we know that the network is missing some of the information that is extracted by the power spectrum. Two questions then arise: (1) how much better will the constraints be if we combine the CNN with the power spectrum, and (2) which part of the Gaussian information, captured by the power spectrum, is missed by the CNN? To answer these questions, we divide the power spectrum into three ranges: the large scale ($100<\ell<400$, 5 bins), the medium scale ($400<\ell<2000$, 6 bins), and the small scale ($2000<\ell<12000$, 7 bins), where $\ell=400$ and $\ell=2000$ corresponds to angular scales of around $30\unit{arcmin}$ and $6\unit{arcmin}$ respectively. We calculate the posterior from the combination of CNN predictions and certain ranges of the power spectrum; the resulting $S_\mathrm{full}$ and $S_\mathrm{fid}$ are shown in Table~\ref{tab:posterior-area}. 

First, we find that adding the full power spectrum to our CNN only improves $S_\mathrm{fid}$ by $10\%$. This indicates that without the influence of baryonic physics, the CNN can extract almost all Gaussian information. Second, adding the power spectrum to our CNN improves $S_\mathrm{full}$ by a factor of 2.2, and $S_\mathrm{full}/S_\mathrm{fid}$ reduces from from 4.70 to 2.41, making the statistics much less affected by the uncertainty in the baryonic physics.

These findings can be understood from two different perspectives: (1) the CNN is less efficient in extracting information from medium and large scales with the presence of baryons, so its constraints can be augmented by the power spectrum; or (2) the degeneracies between the cosmological and baryonic parameters are different from the CNN and from the power spectrum, so combining them enables us to break these cosmology-baryon physics degeneracies and obtain tighter constrains on all six parameters. 

\subsection{Alternative network architectures}
\label{sec:alternative-arch}

\begin{table}[!ht]
\centering
\begin{tabular}{ccccc}
\hline
\multicolumn{2}{c}{Network} & \multicolumn{3}{c}{$\Om-\s8$} \\
$n_\mathrm{ch}$ & $n_\mathrm{layer}$ & $S_\mathrm{full}$ & $S_\mathrm{fid}$ & $S_\mathrm{full}/S_\mathrm{fid}$ \\ \hline
32 & 5  & 1.99 & 0.29 & 6.87 \\
32 & 10 & 1.87 & 0.26 & 7.15 \\
32 & 15 & 1.83 & 0.28 & 6.56 \\
64 & 5  & 1.87 & 0.28 & 6.65 \\
\bf 64 & \bf 10 & \bf 1.62 & \bf 0.28 & \bf 5.85 \\
64 & 15 & 1.73 & 0.29 & 6.08 \\
96 & 5  & 1.79 & 0.26 & 6.77 \\
96 & 10 & 1.90 & 0.27 & 6.99 \\
96 & 15 & 1.66 & 0.26 & 6.29 \\
\multicolumn{2}{c}{Ribli} & 2.63 & 0.36 & 7.38 \\ \hline
\end{tabular}
\caption{The $1\sigma$ areas in $\Om-\s8$ space, assuming a survey with $A_\mathrm{survey}=1500\unit{deg^2}$ and $n_\mathrm{g}=50\unit{arcmin^{-2}}$ and exploring different CNN architectures. The values of $n_\mathrm{ch}$ and $n_\mathrm{layer}$ determine the architecture of the network. ``Ribli'' denotes an architecture derived from the network used in \citet{ribli2019}. Our fiducial network is shown in boldface. }
\label{tab:alternative-arch}
\end{table}

In this study, we have tested several variants of the fiducial network used in \S~\ref{sec:results}. In Table~\ref{tab:alternative-arch}, we list the $1\sigma$ area in $\Om-\s8$ space from the constraints by various networks, spanning different combinations of $n_\mathrm{layer}=5,10,15$ and $n_\mathrm{ch}=32,64,96$, and the network used in \citet{ribli2019} (with the number of outputs in the last layer increased from two to six). The training of all these networks are kept the same, except that the network derived from \citet{ribli2019} was trained with the learning rate schedule described in their article. 

Across all of the networks we tested, the best-performing (in terms of $S_\mathrm{full}$ in $\Om-\s8$) ones are ($n_\mathrm{ch}=64,n_\mathrm{layer}=10$) and ($n_\mathrm{ch}=96,n_\mathrm{layer}=15$), with all others (except \citealt{ribli2019}) performing very close to them (within $\sim20\%$). The networks with $n_\mathrm{ch}=96$ generally perform better in $S_\mathrm{fid}$, but they are similar to $n_\mathrm{ch}=64$ in $S_\mathrm{full}$. The network derived from \citet{ribli2019} performs $\sim50\%$ worse compared to the other networks with or without baryons, but we note that their architecture and learning rate schedule was optimized for predicting $\Om$ and $\s8$ only.

\subsection{Various Learning rates}

\begin{table}[!ht]
\centering
\begin{tabular}{cccc}
\hline
\multirow{2}{*}{Learning rate} & \multicolumn{3}{c}{$\Om-\s8$} \\
 & $S_\mathrm{full}$ & $S_\mathrm{fid}$ & $S_\mathrm{full}/S_\mathrm{fid}$ \\ \hline
0.001 & 2.46 & 0.32 & 7.78 \\
0.002 & 1.84 & 0.31 & 5.86 \\
0.005 & 1.82 & 0.29 & 6.30 \\
\bf 0.01  & \bf 1.62 & \bf 0.28 & \bf 5.85 \\
0.02  & 2.03 & 0.25 & 8.25 \\ \hline
\end{tabular}
\caption{The $1\sigma$ areas in $\Om-\s8$ space ($1500\unit{deg^2}, 50\unit{arcmin^{-2}}$) from the fiducial network, trained with various initial learning rates. The default learning rate, used in prior figures, is shown in boldface. }
\label{tab:learning-rates}
\end{table}

\begin{figure}[!ht]
\centering
\includegraphics[width=6.8cm]{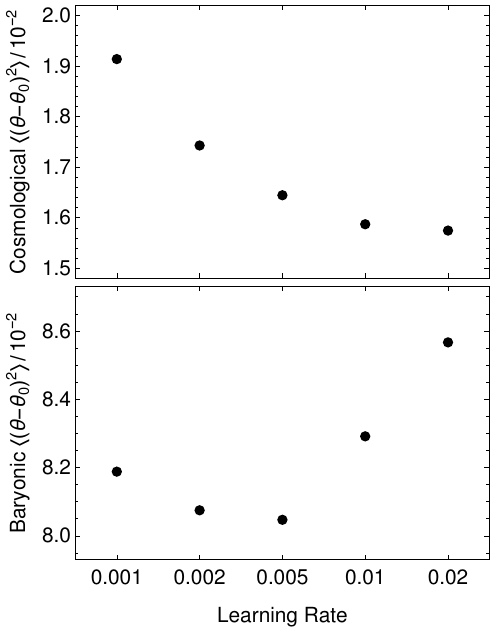}
\caption{Mean square error of the predicted parameterrms as a function of the learning rate during the training of the network. The upper panel shows the average error of the two cosmological parameters, and the lower panel shows that of the four baryonic parameters (re-scaled from their prior to fall in the range $[0,1]$).}
\label{fig:learning-rate-loss}
\end{figure}

The learning rate of the network during the training determines the size of each gradient descent step. Conceptually, this step size controls the speed at which the network learns; the network may take too much time to converge at a too-low learning rate, and it may oscillate around the minimum at a too-high learning rate. To investigate how the learning rate affects the performance of the networks, we scale the learning rate up and down relative to schedule described in \S~\ref{sec:training} (the initial learning rate is $0.01$), and the results are shown in Table~\ref{tab:learning-rates}. We find that as the learning rate increases from 0.001 to 0.02, $S_\mathrm{fid}$ always decreases, while $S_\mathrm{full}$ decreases initially (up to $~0.08$) and increases once the learning rate reaches $0.01$. 

To explain the different behavior of $S_\mathrm{fid}$ and $S_\mathrm{full}$, we evaluate the loss of the network on the cosmological and baryonic parameters separately. As shown in Fig.~\ref{fig:learning-rate-loss}, a larger learning rate improves the predictions on the cosmological parameters, but the predictions on the baryonic parameters degrade when the learning rate is greater than $0.005$. We note that the quality of both sets of predictions affects $S_\mathrm{full}$: the former does it directly, and the latter does it through the degeneracies between cosmological and baryonic parameters. We speculate that an appropriate learning rate (e.g. 0.01 for our task) should balance the quality of prediction on both cosmological and baryonic parameters. A lower learning rate leads to not being able to learn cosmological parameters sufficiently, while a higher learning rate likely misses minor differences between the maps with different baryonic parameters.

\subsection{Dependence on shape noise level}

\begin{table}[!ht]
\centering
\begin{tabular}{ccccc}
\hline
\multirow{2}{*}{Method} & $n_\mathrm{g}$ & \multicolumn{3}{c}{$\Om-\s8$} \\
& $\mathrm{arcmin}^{-2}$ & $S_\mathrm{full}$ & $S_\mathrm{fid}$ & $S_\mathrm{full}/S_\mathrm{fid}$ \\ \hline
CNN                    & 20  & 2.08 & 0.44 & 4.70 \\
CNN                    & 50  & 1.62 & 0.28 & 5.85 \\
CNN                    & 100 & 1.37 & 0.21 & 6.53 \\
Power spectrum         & 20  & 3.45 & 0.93 & 3.71 \\
Power spectrum         & 50  & 2.70 & 0.56 & 4.80 \\
Power spectrum         & 100 & 2.19 & 0.42 & 5.22 \\
CNN$+$Power spectrum & 20  & 0.96 & 0.40 & 2.41 \\
CNN$+$Power spectrum & 50  & 0.71 & 0.23 & 3.09 \\
CNN$+$Power spectrum & 100 & 0.67 & 0.17 & 3.87 \\ \hline
\end{tabular}
\caption{The $1\sigma$ areas in $\Om-\s8$ space for a $1500\unit{deg^2}$ mock survey by the main network and the power spectrum at different noise levels (specified by the galaxy number density $n_\mathrm{g}$).}
\label{tab:noise-levels}
\end{table}

We test the performance of our CNN and the power spectrum using three different sets of maps, with three different noise levels $n_\mathrm{g}=20, 50, 100\unit{arcmin^{-2}}$, where three separate training runs were performed with each of the levels. In all cases, the CNN outperforms the power spectrum by a factor of $\sim1.6$ and $\sim2$ for $S_\mathrm{full}$ and $S_\mathrm{fid}$ respectively. Similar to the power spectrum, the CNN is more affected by the baryons at lower noise levels ($S_\mathrm{full}/S_\mathrm{fid}$ increases from 4.70 to 6.53). This is most likely due to the fact that the extra information available on small scales is mixed more heavily with the baryonic effects and is obscured by these effects. The performance improvement by combining the CNN with the power spectrum is $10-20\%$ for $S_\mathrm{fid}$ and a factor of $\sim2$ for $S_\mathrm{full}$ regardless of the noise level, similar to what we have found in \S~\ref{sec:combine-cnn-ps}.

\subsection{Interpolation errors}

\begin{figure}[!ht]
\centering
\includegraphics[width=8.2cm]{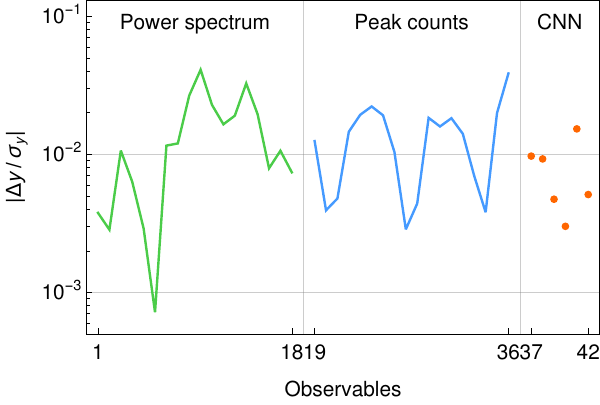}
\caption{The interpolation error of each observable divided by their uncertainties across different convergence map realizations. The errors are calculated at the fiducial cosmological and baryonic parameters with $n_\mathrm{g}=20\unit{arcmin^{-2}}$.}
\label{fig:interpolation}
\end{figure}

To test the accuracy of the interpolation of the observables between discrete cosmologies, we interpolate the observables with all cosmologies expect for the fiducial one and reconstruct the values at the fiducial parameters; we then estimate the interpolation error by the difference between the reconstructed values and the true values. We find that at $n_\mathrm{g}=20\unit{arcmin^{-2}}$, the interpolation errors of the power spectrum and peak counts are $\lesssim5\%$ of the uncertainties (across convergence maps), and those of the CNN is less than $\lesssim2\%$, as are shown in Fig.~\ref{fig:interpolation}. Similar interpolation errors were also found in previous studies; e.g., see \citet{liu2015} for power spectrum and peak counts using a Gaussian process. Assuming a survey with $A_\mathrm{survey}=1500\unit{deg^2}$, the $\chi^2$ associated with the interpolation errors of power spectrum, peak counts, and CNN are 0.33, 0.30, and 0.77, respectively. We note that with the fiducial cosmology present, the interpolation errors around it are expected to be lower than the values estimated above, due to shorter distances to the nearest cosmology. Therefore, we conclude that the process of interpolation is not a significant source of systematic error. 

\section{Conclusions}
\label{sec:conclusions}

In this study, we tested the ability of CNNs to constrain cosmological parameters from WL convergence maps, simultaneously with baryonic effects. We modeled baryonic physics with a baryonic correction model (BCM), applied to the matter distributions in a suite of $N$-body simulations to account for the changes in the density profiles in and near dark matter haloes. We proposed an architecture of the CNNs that consists of convolutions and residual blocks.  We built a fiducial network with 10 residual blocks and 64 channels and trained it on a suite of simulated convergence maps (the training set). We then tested it on a new suite of of simulated maps (the test set) to obtain predictions and posteriors on both the cosmological and the baryonic parameters.

In an HSC-like survey, our CNN outperforms the power spectrum by a factor of 2.10 when the baryonic parameters are fixed and by a factor of 1.66 when we marginalize over baryonic parameters, in terms of the $1\sigma$ area in $\Om-\s8$ space. This shows that the CNN is capable of extracting non-Gaussian information from the convergence maps. When compared to using peak counts, a popular non-Gaussian statistic, the CNN achieves tighter constraints by a factor of 2.83 and 2.14 respectively, which means that the CNN is more efficient in extracting non-Gaussian information than peak counts, even in the face of large uncertainties in baryonic parameters.

When combining the CNN with the power spectrum, the constraint is tighter by a factor of $2.2$ when marginalizing over baryonic parameters compared to using the CNN alone, and the degradation caused by the baryons ($S_\mathrm{full}/S_\mathrm{fid}$) decreases from a factor of 4.70 to 2.41. We find that the performance of the CNN depends on the initial learning rate---a learning rate that is too high or too low degrades the performance. We also investigate the impact of two hyper-parameters $n_\mathrm{ch}$ and $n_\mathrm{layer}$, the number of channels and layers in the network, but find these to be relatively weaker. Similar to the power spectrum, the CNN yields a tighter constraints at lower noise levels, but the impact of baryons is then correspondingly greater. 

It should also be noted that the inclusion of baryonic effects leads to a shift in the cosmological constraint. \citet{fluri2019} have found that when reasonable parameters are taken for the baryonic physics, the constraint on $S_8$ will shift by $\sim5\%$ base on KiDS-450 data. With our results of CNN posteriors at $n_\mathrm{g}=20\unit{arcmin^{-2}}$ ($S_8=0.804\pm0.009$), a similar shift implies a $4\sigma$ bias if baryonic effects are ignored. It means that modelling of baryonic effects (like BCM) is necessary for HSC-like and larger surveys.

We end by noting two significant caveats in this study. First, it remains to be proven that a BCM such as the one considered here can accurately describe the impact of baryonic physics. \citet{schneider2015} and \citet{arico2020} have shown that the BCM can recover the power spectra of various hydrodynamical simulations, but whether the BCM can emulate the non-Gaussian distribution of the matter on small scales, found in these simulations---and ultimately in the real universe---remains to be verified. Since the neural networks are more prone to systematics due to their large numbers of free parameters, the accuracy of the BCM is essential to obtain unbiased constraints. Second, we find that our CNN can distinguish maps with different $M_\mathrm{c}$ and $M_\mathrm{1,0}$, but nearly incapable to determine $\eta$ and $\beta$. This can be mitigated by combining the CNN with the power spectrum or adopting narrower priors based on external observations. For example, \citet{ade2013} suggests $\eta=0.3-0.7$ based on radial profiles of the gas fractions found in galaxy clusters, and \citep{gonzalez2013} suggests $\beta=0.3-0.8$ from the bounded gas fraction--$M_{500}$ relation. The possibility of constraining these parameters by a CNN alone can be explored: increasing the number of the convergence maps and experimenting with other network architectures may facilitate this, and will be addressed in future work. 

\section*{Acknowledgements}

We thank Daniel Hsu for useful discussions and Tomasz Kacprzak for comments that helped us clarify our manuscript. We acknowledge support by NASA ATP grant 80NSSC18K1093, the use of the NSF XSEDE facility Stampede2,  and the Columbia University High-Performance Computing cluster Habanero for the simulations and data analysis used in this study.

\section*{Data availability}

The data underlying this article were accessed from Stampede2. The derived data generated in this research will be shared on reasonable request to the corresponding author.

\bibliography{main}{}
\bibliographystyle{aasjournal}

\end{document}